\documentclass[11pt]{article}
\usepackage{latexsym}

\begin{document}

\begin{center}
{\Large\bf Unravelling Lorentz Covariance  and the Spacetime Formalism \rule{0pt}{13pt}}\par

\bigskip
Reginald T. Cahill \\ 
{\small\it  School of Chemistry, Physics and Earth Sciences, Flinders University,
Adelaide 5001, Australia\rule{0pt}{15pt}}\\
\raisebox{+1pt}{\footnotesize E-mail: Reg.Cahill@flinders.edu.au}\par
\bigskip

{\small\parbox{11cm}{%
We report the discovery of an exact  mapping from Galilean time and space coordinates to Minkowski spacetime coordinates, showing that Lorentz covariance and the spacetime construct are consistent with the existence of a dynamical 3-space, and ``absolute motion".  We illustrate this mapping first with the standard  theory of sound, as vibrations of a medium, which itself may be undergoing fluid motion,  and which is covariant under Galilean 
coordinate transformations.   By introducing a different non-physical class of space and time coordinates it may  be cast into a  form that is covariant under ``Lorentz transformations"  wherein the speed of sound is now the ``invariant speed".  If this latter formalism were taken as fundamental and complete we would be lead to the introduction of a pseudo-Riemannian ``spacetime"  description of sound, with a metric characterised by an  ``invariant speed of sound".  This analysis is an allegory for the development of 20th century physics, but where the Lorentz covariant  Maxwell equations were constructed first, and the Galilean form was later constructed by Hertz, but ignored. It is shown that the Lorentz covariance of the Maxwell   equations only occurs because of the use of  non-physical space and time coordinates. The use of this class of coordinates has confounded 20th century physics, and resulted in  the existence of a ``flowing" dynamical 3-space being overlooked. The discovery of the dynamics of this 3-space has lead to the derivation of an extended gravity theory  as a quantum effect, and confirmed by numerous experiments and observations.
\rule[0pt]{0pt}{0pt}}}\medskip
\end{center}

\setcounter{section}{0}
\setcounter{equation}{0}
\setcounter{figure}{0}
\setcounter{table}{0}

\markboth{R.T.  Cahill, Unravelling Lorentz Covariance  and the Spacetime Formalism}{\thepage}
\markright{R.T.  Cahill,  Unravelling Lorentz Covariance  and the Spacetime Formalism}

\section{Introduction}
It is commonly argued that the manifest success of  Lorentz covariance and the spacetime formalism in Special Relativity (SR) is inconsistent with the anisotropy of the speed of light, and indeed the existence of absolute motion, that is, a detectable motion relative to an actual dynamical 3-space, despite the repeated experimental detection of such effects over, as we now understand, more than 120 years. This apparent incompatibility between a preferred frame, {\it viz} a dynamical 3-space, and  the spacetime formalism is explicitly resolved by the discovery of an exact mapping from Galilean time and space coordinates to Minkowski spacetime coordinates\footnote{See \cite{Mink} and Damour \cite{Damour} for discussion of Minkowski's work.}, showing that Lorentz covariance and the spacetime construct   are indeed consistent with Galilean covariance, but that they suppress any account of an  underlying dynamical 3-space.  

In the neo-Galilean formalism, known also as the Lorentzian interpretation of SR,  length contraction and clock  effects are real effects experienced by objects and clocks in motion relative to an actual 3-space, whereas in the Minkowski-Einstein spacetime formalism these effects are transferred to the metric of the mathematical spacetime, and then appear to be merely perspective effects for different observers.  Experiments, however, have shown that the Galilean space and time coordinates competently describe reality, whereas the Minkowski-Einstein spacetime construct is merely a mathematical artifact, and that various observable phenomena cannot be described by that formalism.   We thus arrive at the dramatic conclusion that the neo-Galilean formalism is the valid description of reality, and that it is a superior more encompassing formalism than the  Minkowski-Einstein formalism in terms of both mathematical clarity and ontology.  

 Physics arrived at the Minkowski-Einstein formalism because of two very significant accidents of history, first that Maxwell's unification of electric and magnetic phenomena failed to build in the possibility of an actual 3-space, for which the speed of light is only $c$ relative to that space, and not relative to observers in general, and 2nd that the first critical test of the Maxwell EM unification by Michelson using interferometry actually suffered  a fundamental design flaw, causing the instrument to be almost 2000 times less sensitive than Michelson had assumed.  A related issue is that the Newtonian theory of gravity used an acceleration field for the description of gravitational phenomena, when a velocity field description would have immediately lead 
to a richer description, and for which notions such as ``dark matter" and ``dark energy" are not needed.

We illustrate the properties of this new  mapping first with the standard  theory of sound, as vibrations of a medium which itself may be undergoing fluid motion,  and which is covariant under Galilean  coordinate transformations, which relate the observations by different observers who may be in motion wrt the fluid and wrt one another.   Here we show that by introducing a different non-physical class of space and time coordinates, essentially  the Minkowski coordinates,  the sound vibration dynamics may be cast into a form that is covariant under ``Lorentz transformations",  wherein the speed of sound is now the invariant speed.  If this latter formalism were taken as fundamental and complete we would be lead to the introduction of a pseudo-Riemannian ``spacetime"  formalism for sound with a metric characterised by the invariant speed of sound, and where ``sound cones" would play a critical role.  

This analysis is an allegory for the development of 20th century physics, but where the Lorentz covariant  Maxwell equations  were constructed first, and the Galilean form was later suggested by Hertz, but ignored. It is shown that the Lorentz covariance of the Maxwell  equations only occurs because of the use of  degenerate non-physical space and time coordinates. The conclusion is that Lorentz covariance and the spacetime formalism are artifacts  of the use of peculiar non-physical space and time coordinates. The use of this class of coordinates has confounded 20th century physics, and lead to the existence of a ``flowing" dynamical 3-space being overlooked. The dynamics of this 3-space, when coupled to the new Schr\"{o}dinger and Dirac equations, has lead to the derivation of an extended gravity theory confirmed by numerous experiments and observations. This analysis also shows that Lorentz symmetry is consistent with the existence of a preferred frame, namely that defined by the dynamical 3-space.  This dynamical 3-space has been repeatedly detected over more than 120 years of experiments, but has always been denied because of the obvious success of the Lorentz covariant formalism, where there the Lorentz transformations are characterised by the so-called invariant speed of light. Einstein's fundamental principle that `the speed of light is invariant"  is not literally true, it is only valid if one uses the non-physical space and time coordinates. 

As with sound waves, the non-invariance  or speed anisotropy of the actual speed of light in vacuum is relatively easy to measure,  and is also relatively large, being approximately 1 part in 1000 when measured on earth, with the direction of the ``flowing space" known since the 1925/26 experiment by Miller \cite{Miller}. Successful direct and sufficiently accurate measurements of the one-way speed of light  have never been made simply because the speed of light is so fast that accurate timing for laboratory-sized speed measurements are not possible. For that reason indirect measurements have always been used. One of the first was the Michelson interferometer.  However a subtlety  always arises for indirect measurements - namely that the anisotropy of the speed of light also affects the operation of the experimental apparatus in ways that have not  always been apparent.  The Michelson interferometer, for example, has a major design flaw that renders it nearly 2000 times less sensitive than believed by Michelson, who used Newtonian physics in calibrating his instrument. It was only in 2002 \cite{MMCK,MMC} that the correct calibration of the Michelson interferometer was derived, and analysis of the non-null fringe shift data from that Michelson-Morley 1887 experiment was analysed and shown to reveal a ``flowing space" with a speed in excess of 300km/s.   The 2002 analysis \cite{MMCK,MMC} showed that the presence of a gas in the Michelson interferometer was a key component of its operation - for in vacuum mode the instrument is totally defective  as a detector of light speed anisotropy.  This is merely because different unrelated effects just happen to cancel when the Michelson interferometer is used in vacuum mode - a simple design flaw that at least Michelson could not have known about.  It so happens that having a gas in the light paths causes this cancellation  to be incomplete. The sensitivity of the instrument varies as $n-1$, where $n$ is the refractive index. For gases this calibration factor is very small - for air at STP $n-1=0.00029$, whereas Michelson, using Newtonian physics, used a calibration coefficient of value 1.   However  if we use optical fibers in place of  air $n-1\approx 0.5$, and the detector is some 2000 times more sensitive, and the use of such detectors has lead to the detailed characterisation of turbulence in the 3-space flow - essentially gravitational waves\footnote{The design flaw of the vacuum-mode Michelson interferometer has been repeated  in the  large and expensive terrestrial gravitational wave detectors such as LIGO, and also in the vacuum-mode resonant cavity interferometers \cite{cavities}. These cavity experiments are based on two mistaken notions:  (i) that a breakdown of Lorentz symmetry is related  to the existence of a preferred frame, and (ii) that vacuum-mode Michelson interferometers can detect a light speed anisotropy associated with such a preferred frame.}. 

There are now four different experimental techniques for detecting light speed anisotropy: (1) gas-mode Michelson interferometer   \cite{Miller,MM,Illingworth,Joos, Jaseja, Munera},  (2) one-way RF speed in coaxial cables \cite{Torr,DeWitte,CahillCoax}, (3) optical fiber interferometer \cite{CahillOF1, CahillOF2}, and (4) doppler-shift effects in earth-flyby of spacecraft \cite{Doppler}. These consistent light-speed anisotropy experiments reveal earth rotation and orbit effects, and sub-mHz gravitational waves. The detection of gravitational wave effects, it now turns out,   dates back to the pioneering work of Michelson and Morley in 1887 \cite{MM}, as discussed in \cite{Review}, and detected again by Miller  \cite{Miller} also using a gas-mode Michelson interferometer, and by Torr and Kolen \cite{Torr}, DeWitte \cite{DeWitte}  and Cahill \cite{CahillCoax}  using RF waves in coaxial cables, and by Cahill \cite{CahillOF1} and Cahill and Stokes \cite{CahillOF2} using an optical-fiber interferometer design, and also present in the spacecraft flyby doppler shifts \cite{Doppler}.

\section{Sound Wave Galilean Covariant Formalism}
Let us first use the example of sound waves to discuss the mapping from Galilean space and time coordinates to Minkowski-Einstein spacetime coordinates - as in this case the underlying physics is well understood. 
The standard  formulation for sound waves in a moving fluid is
\begin{equation}
\left(\frac{\partial}{\partial t} +{\bf v}({\bf r},t).\nabla \right)^2\phi({\bf r},t)=c^2\nabla^2\phi({\bf r},t).
\label{eqn:sound1}\end{equation}
where $\nabla=\{\frac{\partial}{\partial x},\frac{\partial}{\partial y},\frac{\partial}{\partial z}\}$. The physical time coordinate $t$ and  Euclidean  space coordinates ${\bf r}=\{x,y,z\}$  are used by an observer  $O$ to label the readings of a clock and the location in space where the ``wind" or ``fluid flow"  has velocity ${\bf v}({\bf r},t)$, and small pressure variations $\phi({\bf r},t)$, relative to the background pressure.  Clearly the ``fluid flow" and ``pressure fluctuations" are different aspects of the same underlying phenomena - namely 
the dynamics of some macroscopic system of  atoms and/or molecules, but separated into very low frequency effects, - the flow, and high frequency effects,  - the sound waves.  The dynamics for the flow velocity ${\bf v}({\bf r},t)$ is not discussed here.  As well the symbol $c$ is the speed of sound waves relative to the fluid. In (\ref{eqn:sound1}) the coordinates  $\{t,x,y,z\}$ ensure that the dynamical flow ${\bf v}$  is correctly related to the pressure fluctuation $\phi$, at the same time and space. Whenever we separate some unified phenomenon into two or more   related phenomena we must introduce a ``coordinate system" that keeps track of the connection.
To demonstrate this we find plane-wave solutions of (\ref{eqn:sound1}) for the case where the fluid flow velocity is time and space independent, {\it viz} uniform,
\begin{equation}
\phi({\bf r},t)=A\sin({\bf k}.{\bf r}-\omega t)
\label{eqn:sound2}\end{equation}
\begin{equation}
\omega({\bf k},{\bf v})=c|\vec{{\bf k}}| +{\bf v}.{\bf k}
\label{eqn:sound3}\end{equation}
The sound wave group velocity is then
\vspace{-1.5mm}
\begin{equation}
{\bf v}_g=\vec{\nabla}_k\omega({\bf k},{\bf v})=c\hat{\bf k}+{\bf v}
\label{eqn:groupv}\end{equation}
and we see that the wave has velocity ${\bf v}_g$ relative to the observer, with the fluid flowing at velocity  ${\bf v}$ also relative  to the observer, and so the speed of sound is $c$ in direction $\hat{\bf k}$ relative to the fluid itself. This corresponds to a well known effect, namely that sound travels slower up-wind than down-wind. This ``sound speed anisotropy" effect can be measured by means of one-way sound travel times, or indirectly by means of   doppler shifts for   sound waves reflected from a distant object separated by a known distance from the observer.

Next consider two observers, $O$ and $O^\prime$, in relative motion. Then the physical time and space coordinates of each are related by the Galilean transformation
\vspace{-3mm}
\begin{eqnarray}
t^\prime&=&t, \nonumber \\
x^\prime&=&x-Vt,  \mbox{\ \ \  }y^\prime=y, \mbox{\ \ \  }z^\prime=z.
\label{eqn:GT}\end{eqnarray}
We have taken the simplest case where $V$ is the relative speed of the two observers   in their common $x$ directions. Then the derivatives  are related by
\begin{eqnarray}
\frac{\partial}{\partial t}&=&\frac{\partial}{\partial t^\prime}-V\frac{\partial}{\partial x}, \nonumber \\
\frac{\partial}{\partial x}&=&\frac{\partial}{\partial x^\prime}, \frac{\partial}{\partial y}=\frac{\partial}{\partial y^\prime}, \frac{\partial}{\partial z}=\frac{\partial}{\partial z^\prime}.
\label{eqn:derivs}\end{eqnarray}
Then (\ref{eqn:sound1}) becomes for the 2nd observer, with $v^\prime=v-V$,
\begin{equation}
\left(\frac{\partial}{\partial t^\prime} +{\bf v}^\prime({\bf r}^\prime,t^\prime).\nabla^\prime \right)^2\phi(^\prime{\bf r}^\prime,t^\prime)=c^2\nabla^{\prime 2}\phi(^\prime{\bf r}^\prime,t^\prime).
\label{eqn:sound4}\end{equation}
For  sound waves $\phi(^\prime{\bf r}^\prime,t^\prime)=\phi({\bf r},t)$.
If the flow velocity  ${\bf v}({\bf r},t)$ is not uniform then we obtain refraction effects for the sound waves. Only for an observer at rest in a time independent and uniform  fluid does $v^\prime$  disappear from (\ref{eqn:sound4}).

\section{Sound Wave Lorentz Covariant Formalism}
The above Galilean formalism  for sound waves is well known and uses physically sensible choices for the time and space coordinates. Of course we could choose to use spherical or cylindrical space coordinates if we so desired. This would cause no confusion.  However we could also  choose to use a new class of time and space coordinates, indicated by upper-case symbols $T,X,Y,Z$, that mixes the above time and space coordinates.  One such new class of coordinates is
\begin{eqnarray}
T&=&\gamma(v)\left((1-\frac{v^2}{c^2})t+\frac{v x}{c^2}\right), \nonumber \\
X&=&\gamma(v)x; \mbox{\ \ \  }Y=y; \mbox{\ \ \  }Z=z,
\label{eqn:GtoL}\end{eqnarray}
where $\gamma(v)=1/\sqrt{1-v^2/c^2}$. Note that this is not a Lorentz transformation. The transformations for the derivatives are then found to be
\vspace{-1mm}
\begin{eqnarray}
\frac{\partial}{\partial t}\!\!\!\!\!\!&=&\!\!\!\!\!\!\gamma(v)(1-\frac{v^2}{c^2})\frac{\partial}{\partial T}, \nonumber \\
\frac{\partial}{\partial x}\!\!\!\!\!\!&=&\!\!\!\!\!\!\gamma(v)\left(\frac{v}{c^2}\frac{\partial}{\partial T}+\frac{\partial}{\partial X}\right), \frac{\partial}{\partial y}=\frac{\partial}{\partial Y}, \frac{\partial}{\partial z}=\frac{\partial}{\partial Z}.
\label{eqn:sound5}\end{eqnarray}
We define $\overline{\nabla}=\{\frac{\partial}{\partial X},\frac{\partial}{\partial Y},\frac{\partial}{\partial Z}\}$. Then (\ref{eqn:sound1}) becomes, for uniform $v$,
\begin{equation}
\left(\frac{\partial}{\partial T}  \right)^2\overline{\phi}({\bf R},T)=c^2\overline{\nabla}^2\overline{\phi}({\bf R},T).
\label{eqn:sound6}\end{equation}
with ${\bf R}=\{X,Y,Z\}$ and $\overline{\phi}({\bf R},T)=\phi({\bf r},t)$. This is a remarkable result. In the new class of coordinates the dynamical equation no longer contains the flow velocity  $\bf v$ - it has been mapped out of the dynamics. Eqn.(\ref{eqn:sound6}) is now covariant under Lorentz transformations\footnote{Lorentz did not construct the ``Lorentz transformation" - and this nomenclature is very misleading as Lorentz held to a different interpretation of the so-called relativistic effects.}, 
\vspace{-3mm}
\begin{eqnarray}
T^\prime&=&\gamma(V)\left(T+\frac{VX}{c^2}\right), \nonumber \\
X^\prime&=&\gamma(V)(X-VT), \mbox{\ \ \  }Y^\prime=Y,  \mbox{\ \ \  }Z^\prime=Z,
\label{eqn:LT}\end{eqnarray}
where we have taken the simplest case, and where $V$ is a measure of the  relative speed of the two observers   in their common $X$ directions.

There is now no reference to the underlying flowing fluid system  - for an observer using this class of space and time coordinates the speed of sound  relative to the observer is  always $c$ and so invariant - there will be no sound speed  anisotropy.  We could also introduce a ``spacetime" construct with pseudo-Riemannian metric $ds^2=c^2dT^2- d{\bf R}^2$, and sound cones along which $ds^2=0$.  As well pairs of spacetime events could be classified into either time-like or space-like, with the time ordering of spacelike events not being uniquely defined.

However this sound-speed invariance  is purely an artifact of the non-physical space and time coordinates introduced in (\ref{eqn:GtoL}). The non-physical nature of this inferred ``invariance" would have been easily exposed by doing measurements of the speed of sound in different directions.  However in a bizarre imaginary  world the Lorentz-covariant sound formalism could have been discovered first, and the spacetime formalism might have been developed and become an entrenched belief system.  If later experiments had revealed that the speed of sound was actually anisotropic then the experimentalist involved  might have been applauded, or, even more bizarrely, their discoveries denied and suppressed, and further experiments stopped by various means. The overwhelming evidence is that this bizarre possibility is precisely what happened for electromagnetics, for Maxwell essentially introduced the Lorentz covariant electromagnetism formalism, and experiments that  detected the light speed anisotropy have been attacked and any discussion and analysis of the data suppressed - only experiments that failed, for various reasons, have been  reported in mainstream journals.  
 
\section{Dynamical 3-Space Theory}
Here we briefly review the dynamics of the  3-space that is the analogue of the ``flowing fluid" in the sound allegory. For zero vorticity we have \cite{Book,Review,QC} 
\begin{eqnarray}
\nabla.\left(\frac{\partial {\bf v} }{\partial t}+ ({\bf v}.{\bf \nabla}){\bf v}\right)+
\frac{\alpha}{8}\left((tr D)^2 -tr(D^2)\right)=-4\pi G\rho,  \nonumber 
\end{eqnarray}
\vspace{-6mm}
\begin{eqnarray}
 \nabla\times {\bf v}={\bf 0},  \mbox{\  \  \   }
D_{ij}=\frac{1}{2}\left(\frac{\partial v_i}{\partial x_j}+ \frac{\partial v_j}{\partial x_i}\right),
\label{eqn:S1}\end{eqnarray} 
where $\rho({\bf r},t)$ is the matter and EM energy densities expressed as an effective matter density.  Experiment and astrophysical data has shown that $\alpha\approx1/137$ is  the fine structure constant to within observational errors  \cite{Book, Schrod,Review,QC}.  For a quantum system with mass
$m$ the Schr\"{o}dinger equation must be  generalised   \cite{Schrod} with the new terms required to maintain that the motion is intrinsically wrt to the 3-space and that the time evolution is unitary
\begin{equation}
i\hbar\frac{\partial \psi({\bf r},t)}{\partial t}  =-\frac{\hbar^2}{2m}\nabla^2\psi({\bf r},t)-i\hbar\left({\bf
v}.\nabla+\frac{1}{2}\nabla.{\bf v}\right) \psi({\bf r},t).
\label{eqn:Schrod}\end{equation}
The space and time coordinates $\{t,x,y,z\}$ in (\ref{eqn:S1}) and (\ref{eqn:Schrod}) ensure that  the separation of a deeper and unified process into different classes of phenomena - here a dynamical 3-space and a quantum system, is properly tracked and connected. As well the same coordinates may be used by an observer to also track the different phenomena.  However it is important to realise that these coordinates have no ontological significance - they are not real. Nevertheless it is imperative not to use a degenerate system of coordinates that suppresses the description of actual phenomena. The velocities ${\bf v}$ have no ontological or absolute meaning relative to this coordinate system - that is in fact how one arrives at the form in  (\ref{eqn:S1}), and so the ``flow" is always relative to the internal dynamics of the 3-space. So now this is different to the example of sound waves.  

A wave packet propagation analysis gives  the acceleration induced by wave refraction to be \cite{Schrod}
\begin{equation}
{\bf g}=\frac{\partial{\bf v}}{\partial t}+({\bf v}.\nabla){\bf v}+
(\nabla\times{\bf v})\times{\bf v}_R,
\label{eqn:acceln}\end{equation}
 \vspace{-4mm}
\begin{equation}
{\bf v}_R({\bf r}_0(t),t) ={\bf v}_0(t) - {\bf v}({\bf r}_0(t),t),
\end{equation}
which is the velocity of the wave packet relative to the 3-space, and where ${\bf v}_O$ and ${\bf r}_O$ are the velocity and position relative to the observer, and the last term in (\ref{eqn:acceln}) generates the Lense-Thirring effect as a vorticity driven effect.  Together (\ref{eqn:S1}) and (\ref{eqn:acceln}) amount to the derivation of gravity as a quantum effect,  explaining  both the  equivalence principle ($\bf g$ in (\ref{eqn:acceln}) is independent of $m$) and the Lense-Thirring effect. Overall we see, on ignoring vorticity effects, that
\begin{equation}
\nabla.{\bf g}=-4\pi G\rho-\frac{\alpha}{8}\left((tr D)^2 -tr(D^2)\right),
\label{eqn:NGplus}\end{equation}
which is Newtonian gravity but with the extra dynamical term whose strength is given by $\alpha$. This new dynamical effect explains the spiral galaxy flat rotation curves  (and so doing away with the need for ``dark matter"), the bore hole $g$ anomalies,  the black hole ``mass spectrum". Eqn.(\ref{eqn:S1}), even when $\rho=0$, has an expanding universe Hubble solution that fits the recent supernovae data in a parameter-free manner without requiring ``dark matter" nor "dark energy", and without the accelerating expansion artifact \cite{QC}. However (\ref{eqn:NGplus}) cannot be entirely expressed in terms  of ${\bf g}$ because the fundamental dynamical variable is $\bf v$. The role of  (\ref{eqn:NGplus}) is to reveal that if we analyse gravitational phenomena we will usually find that the matter density $\rho$ is insufficient to account for the observed ${\bf g}$. Until recently this failure of Newtonian gravity has been explained away as being caused by some unknown and undetected ``dark matter" density.  Eqn.(\ref{eqn:NGplus}) shows that to the contrary it is a dynamical property of 3-space itself.

Another common misunderstanding is that the success of the Direc equation implies that a preferred frame cannot exist.  This belief is again easily demolished.  The generalised Dirac equation which uses the Galilean class of space-time coordinates is
\begin{equation}
i\hbar\frac{\partial \psi}{\partial t}=-i\hbar\left(  c{\vec{ \alpha.}}\nabla + {\bf
v}.\nabla+\frac{1}{2}\nabla.{\bf v}  \right)\psi+\beta m c^2\psi,
\label{eqn:Dirac1}\end{equation}
where $\vec{\alpha}$ and $\beta$ are the usual Dirac matrices.  This equation shows that the Dirac spinor propagates wrt to the 3-space, and that there are dynamical effects associated with that that are not in the generalised Schr\"{o}dinger equation (\ref{eqn:Schrod}).  As shown elsewhere (\ref{eqn:Dirac1}) gives rise to relativistic gravitational effects\footnote{Meaning when an object has speed comparable to $c$ wrt the 3-space.}, that go beyond those in (\ref{eqn:acceln}).

\section{Galilean Covariant Electromagnetic Theory}
Hertz in 1890  \cite{Hertz} noted that Maxwell had overlooked  the velocity field that accompanies time derivatives, as in (\ref{eqn:sound1}),
 and presented an improved formalism, and the minimal source-free form  is
 \vspace{-5mm}
\begin{eqnarray}
\mu\left(\frac{\partial }{\partial t}+{\bf v.\nabla }\right) {\bf H}&=&-\nabla \times {\bf E},\nonumber \\
\epsilon\left(\frac{\partial }{\partial t}+{\bf v.\nabla }\right){\bf E}&=&+\nabla \times {\bf H},  \nonumber \\
\nabla.{\bf H}={\bf 0},  & & \!\!\!\!\!\!\!\!\!\! \nabla.{\bf E}={\bf 0},
\label{eqn:Max1}\end{eqnarray}
with ${\bf v}({\bf r},t)$  being the dynamical 3-space velocity field as measured\footnote {Earth based light speed anisotropy experiments show that $v$ has  value $\approx$ 420$\pm$30 kms in a known direction \cite{Review}, and is not to be confused with the CMB velocity.} by some observer using time and space coordinates $\{t,x,y,z\}$, although Hertz did not consider a time and space dependent ${\bf v}$. Again for uniform and time-independent $\bf v$  (\ref{eqn:Max1}) has plane wave solutions 
\begin{equation}
{\bf E}({\bf r},t)={\bf E}_0e^{i({\bf k}.{\bf r}-\omega t)}, \mbox{\ \ \ \  } {\bf H}({\bf r},t)={\bf H}_0e^{i({\bf k}.{\bf r}-\omega t)}
\label{eqn:pw}\end{equation}
\begin{equation}
\omega({\bf k},{\bf v})=c|\vec{{\bf k}}| +{\bf v}.{\bf k} \mbox{ \ \ \  where \ \ \  } c=1/\sqrt{\mu\epsilon}.
\label{eqn:omega}\end{equation}
Then the EM group velocity is
 \vspace{-2mm}
\begin{equation}
{\bf v}_{EM}=\vec{\nabla}_k\omega({\bf k},{\bf v})=c\hat{\bf k}+{\bf v}.
\label{eqn:groupv2}\end{equation} So, like the analogy of sound,  the velocity of EM radiation ${\bf v}_{EM}$ has magnitude  $c$ only with respect to the 3-space, and in general not with respect to the observer if the observer is moving through that 3-space, as experiment has indicated again and again, as discussed above. 
Eqns.(\ref{eqn:Max1}) give, for uniform $\bf v$,
 \vspace{-6mm}
\begin{eqnarray}
\left(\frac{\partial }{\partial t}+{\bf v.\nabla }\right)^2 {\bf E}&=&c^2\nabla^2{\bf E},\nonumber \\
\left(\frac{\partial }{\partial t}+{\bf v.\nabla }\right)^2 {\bf H}&=&c^2\nabla^2{\bf H}.
\label{eqn:Max2}\end{eqnarray}
on using the identity $\nabla\times(\nabla\times {\bf E})=-\nabla^2{\bf E}+\nabla(\nabla.{\bf E})$ and $\nabla.{\bf E}=0$, and similarly for the ${\bf H}$ field.  Transforming to the Minkowski-Einstein $T,X,Y,Z$ coordinates using (\ref{eqn:GtoL}) and (\ref{eqn:sound5})  we obtain the form of the source-free ``standard" Maxwell equations
 \vspace{-2mm}
\begin{equation}
\frac{\partial^2 {\bf E}}{\partial T^2} =c^2\overline{\nabla}^2{\bf E},\mbox{\ \ \  }\frac{\partial^2 {\bf H}}{\partial T^2} =c^2\overline{\nabla}^2{\bf H}
\label{eqn:Max3}\end{equation}
which is again covariant under Lorentz transformation  (\ref{eqn:LT}). It is important to emphasize that the transformation from the Galilean covariant Hertz-Maxwell equations (\ref{eqn:Max1}) to the Lorentz covariant Maxwell equations (\ref{eqn:Max3})  is {\it exact}. It is usually argued that the Galilean transformations   (\ref{eqn:GT}) are the non-relativistic limit of the Lorentz transformations   (\ref{eqn:LT}). While this is technically so, as seen by taking the limit $v/c\rightarrow 0$, this  misses the key point that they are related by the new mapping in (\ref{eqn:GtoL}).  Also we note that for the Galilean space-time class  the speed of light is anisotropic, while it is isotropic for the Minkowski-Einstein  space-time class.  It is only experiment that can decide which of the two classes of coordinates is the more valid space-time coordinate system.   As noted above, and since 1887,  experiments have detected that the speed of light is indeed anisotropic. 

Again when using the Minkowski-Einstein coordinates there is now no reference to the underlying dynamical 3-space system  - for an observer using this class of space and time coordinates the speed of light  relative to the observer is  always $c$ and so invariant.  We could then be tricked into  introducing a ``spacetime" construct with pseudo-Riemannian metric $ds^2=c^2dT^2- d{\bf R}^2$, and light cones along which $ds^2=0$.  As well pairs of spacetime events could be classified into either time-like or space-like, with the time ordering of spacelike events not being uniquely defined.  This loss of the notion of simultaneity is merely a consequence of the degenerate nature of the Minkowski-Einstein spacetime coordinates.  This has confounded  progress in physics for more than a century.

Hence the Minkowski-Einstein  space-time coordinates are degenerate in that they map out the existence of the dynamical 3-space.  So the development of 20th century physics has been misled  by two immensely  significant ``accidents", 1st that Maxwell failed to include the velocity ${\bf v}$, and the 2nd  that the Michelson interferometer in gas-mode is some 2000 times less sensitive than Michelson had assumed,  and that the observed fringe shifts actually indicate a large value for $v$ in excess of 300km/s.  These two accidents stopped physics from discovering the existence of a dynamical 3-space, until recently, and that the dynamical 3-space displays wave effects.  Also again this transformation  between the two classes of space-time coordinates explicitly demonstrates that ``Lorentz covariance" coexists with a preferred frame, contrary to the aims of the experiments in  \cite{cavities}. Furthermore vacuum-mode Michelson interferometers, such as the vacuum cavity resonators,  cannot even detect the long-standing light speed anisotropy.  
We can apply the inverse mapping, from the Minkowski-Einstein class to the Galilean class of coordinates,  but in doing so we have lost the value of the velocity field. In this sense the Minkowski-Einstein class is degenerate -it cannot be used to analyse light speed anisotropy experiments for example.

\section{Conclusions}

We have reported herein the discovery of an exact and invertible mapping from Galilean time and space coordinates to Minkowski-Einstein spacetime coordinates. This mapping removes the effects of the velocity of the dynamical 3-space relative to an observer, and so in this sense the Minkowski-Einstein  coordinates are degenerate - they stop the usual Special Relativity formalism from being able to say anything about the
existence of a preferred frame,  a real 3-space, and from describing experiments that have detected light speed anisotropy. The Minkowski-Einstein formalism has nevertheless  has been very successful in describing other effects.  The spacetime formalism, with its spacetime metric and Lorentz covariance, is really an artifact of the degenerate Minkowski-Einstein coordinates, and we have shown how one may unravel these mathematical artifacts, and display the underlying dynamics..   The new mapping shows that relativistic effects are  caused by motion relative to an actual 3-space - and which has been observed for more than 120 years.   This was Lorentz's proposition.   The belief that spacetime actually described reality has lead to numerous misconceptions about the nature of space and time.  These are distinct phenomena, and are not fused into some 4-dimensional entity.  Indeed time is now seen to have a cosmic significance, and that all observers can measure that time - for by measuring their local absolute speed relative to their local 3-space they can correct the ticking rate of their clocks to remove the local time dilation effect, and so arrive at a measure of the ticking rate of cosmic time\footnote{Uncorrected Earth-based clocks lose approximately 0.085s per day compared to cosmic time, because $v \approx$ 420 km/s.}.  This changes completely how we might consider modelling deeper reality - one such proposition is   {\it Process Physics} \cite{Book,Review,QC}

The Special Relativity formalism asserts that only relative descriptions of phenomena  between two or more observers have any meaning. In fact we now understand that all effects are dynamically and observationally relative to an ontologically real, that is, detectable dynamical 3-space. Ironically this situation has always been known as an ``absolute effect".  The most extraordinary outcome of recent discoveries is that  a dynamical 3-space exists,  and that from the beginning of Physics this has been missed - that a most fundamental aspect of reality has been completely overlooked, and that furthermore all  experimental evidence for this has been suppressed by mainstream physics,

\end{document}